\documentclass[11pt]{article}
\usepackage{graphicx}

\newsavebox{\hflrar}
\sbox{\hflrar}{\makebox[0pt][l]
{${\scriptstyle \leftharpoonup}$}{${\scriptstyle \rightharpoonup}$}}

\def \to {\rightarrow}

\begin{document}

\begin{center}
{\Large\bf Gauge Invariance and $k_T$-Factorization of Exclusive Processes}
\vskip 10mm
F. Feng$^1$, J.P. Ma$^1$ and Q. Wang$^{2}$    \\
{\small {\it $^1$ Institute of Theoretical Physics, Academia Sinica,
Beijing 100080, China }} \\
{\small {\it $^2$ Department of Physics and Institute of Theoretical
Physics, Nanjing Normal University, Nanjing, Jiangsu 210097, P.R.China}} \\
\end{center}
\vskip 10mm

\begin{abstract}
In the $k_T$-factorization for exclusive processes, the nontrivial $k_T$-dependence
of perturbative coefficients, or hard parts, is obtained by taking off-shell partons. This
brings up the question of whether the $k_T$-factorization is gauge invariant.
We study the $k_T$-factorization for the case $\pi \gamma^* \to \gamma$ at one-loop
in a general covariant gauge. Our results show that the hard part
contains a light-cone singularity that is absent in the Feynman gauge, which indicates that the $k_T$-factorization is {\it not} gauge invariant.
These divergent contributions come from the $k_T$-dependent wave function
of $\pi$ and are not related to a special process.
Because of this fact the $k_T$-factorization for any process is not
gauge invariant and is violated. Our study also indicates
that the $k_T$-factorization used widely for exclusive $B$-decays
is not gauge invariant and is violated.
\vskip 5mm \noindent
\end{abstract}
\vskip 1cm
\par
When large momentum transfers happen in exclusive processes, one can employ
perturbation theory of QCD thanks to its asymptotic freedom.
However, a pre-condition for using the perturbation theory is
to consistently factorize perturbative effects from nonperturbative effects
in amplitudes of exclusive processes.
It has been proposed long time ago that the factorization of amplitudes can be made
by expanding amplitudes in the inverse of the large momentum transfers\cite{BL,CZrep}.
The expansion corresponds to the twist expansion of QCD operators characterizing
nonperturbative effects.
The leading term can be
factorized as a convolution of a hard part and light-cone wave
functions of hadrons. The light-cone wave functions are defined with
QCD operators, and the hard part describes hard scattering of
partons at short distances. In this factorization the transverse
momenta of partons in parent hadrons are also expanded in the hard
scattering part, and they are neglected in the leading twist. Because of this, the light-cone wave functions depend only on the
longitudinal momentum fractions carried by partons, not on
transverse momenta of partons. This is the so-called collinear
factorization.
\par
Later, $k_T$-factorization was proposed\cite{LS}, in which the
transverse momenta, neglected in collinear factorization, are taken
into account. The effects of the  transverse-momentum of partons in
hadrons are described by wave functions and those in the hard
scattering are described by a hard part which depends on $k_T$. The
$k_T$-factorization has been widely used in exclusive $B$-decays, a
partial list of references can be found in \cite{KTB1,KTB2}, where
it is often called the perturbative QCD(pQCD) approach. The
factorization has some advantages in that it includes some higher
twist effects and re-sums the Sudakov logarithms (see e.g.,
\cite{KTB1,KTB2} and references therein). However, the factorization
has not been examined extensively as the  collinear factorization.
It is possible that the $k_T$-factorization is violated. It should be
noted that the nontrivial $k_T$-dependence of the hard part at the
tree-level is obtained with off-shell partons entering hard
scattering. This brings up the question if the hard part is gauge
invariant. Recently, the $k_T$-factorization has been studied at the
one-loop level for $\pi^0 +\gamma^* \to \gamma$\cite{NLi}. The study
is done with the Feynman gauge and it is claimed that the hard part
with the off-shell parton is gauge invariant\cite{NLi}. In this
letter we examine the $k_T$-factorization of the process in a
general covariant gauge. It turns out that the hard part obtained
with off-shell partons is {\it not} gauge invariant. Moreover, the
hard part in the general covariant gauge contains special
singularities called light-cone singularities, although the hard
part in the Feynman gauge is finite at one-loop according to ref.
\cite{NLi}. The contributions of the singularities come from the
wave function. This indicates that the $k_T$-factorization is not
gauge invariant and is, in general, violated.
\par
We consider the process $\pi^0(P) +\gamma^*\to  \gamma (p)$.
We will use the  light-cone coordinate system, in which a
vector $a^\mu$ is expressed as
$a^\mu = (a^+, a^-, \vec a_\perp) =
((a^0+a^3)/\sqrt{2}, (a^0-a^3)/\sqrt{2}, a^1, a^2)$ and $a_\perp^2
=(a^1)^2+(a^2)^2$.
We take a frame in which $\pi^0$ has the momentum
$P^\mu =(P^+, P^-,0,0)$ with $P^+ \gg P^-$ and the outgoing photon has $p^\mu =(0,p^-,0,0)$.
The relevant form factor and its $k_T$ factorization\cite{NLi} are:
\begin{eqnarray}
&& \langle \gamma (p, \epsilon^*)\vert \bar q\gamma^\mu q \vert \pi^0(P)\rangle
= i \varepsilon^{\mu\nu\rho\sigma} \epsilon^*_{\nu} P_{\rho} p_{\sigma}
F(Q^2),\ \ \  Q^2 = -(P-p)^2,
\nonumber\\
&& F(Q^2 ) = \int dx d^2 k_T \phi (x, k_T) H (x, k_T)\left ( 1 + {\mathcal O} (Q^{-2}) \right ),
\end{eqnarray}
where $\phi$ is the wave function defined below, and
$H$ is the hard part. If the $k_T$-factorization holds,
the hard part $H$ can be calculated safely with perturbative QCD.
We introduce a vector $u^\mu=(u^+,u^-,0,0)$ such that
the wave function for $\pi^0$ can be defined
in the limit $u^+ << u^-$\cite{NLi,LiLi,MW1}:
\begin{eqnarray}
\phi(x, k_T,\zeta, \mu) &=& \ \int \frac{ d z^- }{2\pi}
  \frac {d^2 z_\perp}{(2\pi )^2}  e^{ik^+z^- - i \vec z_\perp\cdot \vec
k_T}
\langle 0 \vert \bar q(0) L_u^\dagger (\infty, 0)
  \gamma^+ \gamma_5 L_u (\infty,z) q(z) \vert \pi^0(P) \rangle\vert_{z^+=0},
\nonumber\\
   k^+ &=& xP^+, \ \ \ \ \  \zeta^2 = \frac{2 u^- (P^+)^2}{u^+}\approx
\frac{ 4 (u\cdot P)^2}{u^2},
\label{def}
\end{eqnarray}
where $q(x) $ is the light-quark field.
$L_u$ is the gauge link in the direction $u$:
\begin{equation}
L_u (\infty, z) = P \exp \left ( -i g_s \int_{0} ^{\infty} d\lambda
     u\cdot G (\lambda u+ z ) \right ) .
\end{equation}
\par
The hard part $H$ is obtained by replacing the hadron state with a parton state.
With the parton state one can calculate the wave function and the form factor, and hence can determine
the hard part. We replace $\pi^0$ with the partonic state $\vert q(k_1), \bar
q(k_{2})\rangle$ with
the momenta given as
\begin{equation}
k_1^\mu =(k_1^+, k_1^-,\vec k_{1\perp}), \ \ \ \  k_2^\mu
=(k_2^+, k_2^-, -\vec k_{1\perp}), \ \ \ \ \ \
k_1^+ = x_0 P^+, \ \ \ \  k_{2}^+ = (1-x_0) P^+ = \bar x_0 P^+.
\end{equation}
If we take the partons on-shell, we have the wave function at tree-level:
\begin{equation}
\phi^{(0)} (x,k_T,\zeta) =  \delta (x-x_0)
     \delta^2(\vec k_T -\vec k_{1\perp})\bar
v(k_{2}) \gamma^+ \gamma_5 u(k_1) /P^+.
\end{equation}
For the same parton state, the form factor receives
contributions from the diagrams in Fig.1.
Through a simple calculation one can determine the hard part at tree-level as:
\begin{equation}
  Q^2 H^{(0)} = \frac{1}{x} + \frac{1} {\bar x}, \ \ \ \  \bar x =1-x.
\end{equation}
\par
\begin{figure}[hbt]
\begin{center}
\includegraphics[width=6cm]{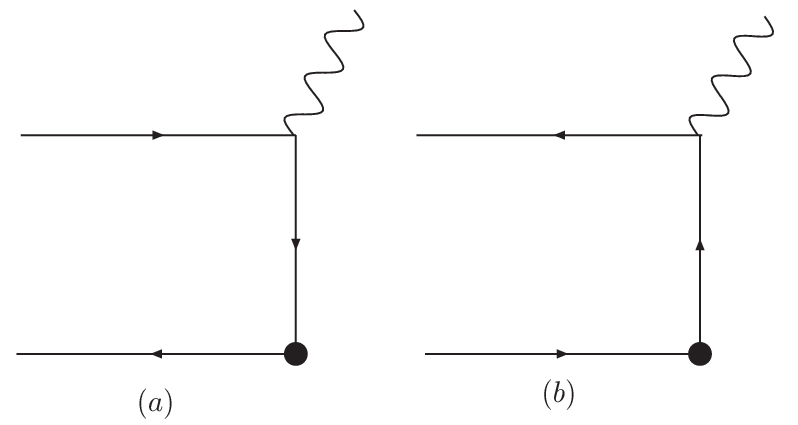}
\end{center}
\caption{The tree-level diagrams of the form factor. }
\label{Feynman-dg1}
\end{figure}
\par
With the on-shell parton, one obtains the hard part which does not depend on $k_T$,
although the partons entering the hard scattering have nonzero transverse momenta.
The dependence can be obtained if one takes the parton state with off-shell partons.
Following the $k_T$-factorization illustrated in ref. \cite{NLi},
we take partons off-shell with the momenta:
\begin{equation}
k_1^\mu =(k_1^+, 0,\vec k_{1\perp}), \ \ \ \  k_2^\mu
=(k_2^+, 0, -\vec k_{1\perp}),
\end{equation}
and replace the product of spinors by:
\begin{equation}
   u (k_1 ) \bar v(k_2) \to \gamma_5 \gamma^-,
\end{equation}
to pick up the leading twist contributions. With the off-shell partons
one indeed gets the hard part depending on $k_T$:
\begin{equation}
   H^{(0)} = \frac{1}{x Q^2 + k^2_T} + \frac{1} {\bar x Q^2 + k^2_T},
\end{equation}
where the first and the second term come from Fig. 1a and Fig. 1b, respectively.
\par
In general, the quantities calculated with off-shell partons are not gauge invariant. The one-loop
hard part extracted from
\begin{equation}
  H^{(1)} \otimes \phi^{(0)} = F^{(1)} - H^{(0)} \otimes \phi^{(1)}
\end{equation}
can not be expected to be gauge invariant. In ref. \cite{NLi}, $H^{(1)}$ is calculated in the Feynman gauge
and it is claimed that $H^{(1)}$ is gauge invariant. In this letter we will study
$H^{(1)}$ in a covariant gauge. Because of the symmetry of charge-conjugation we only
need to consider the one-loop correction of Fig.1a and its factorization.
\par
\begin{figure}[hbt]
\begin{center}
\includegraphics[width=10cm]{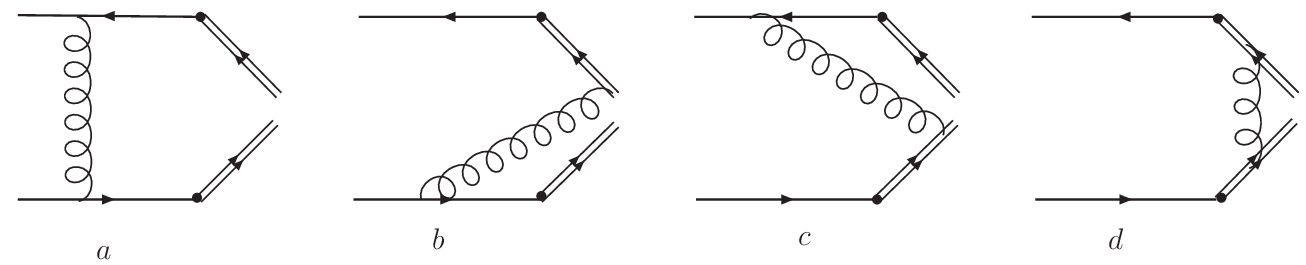}
\end{center}
\caption{The one-loop diagrams of the wave function. }
\label{Feynman-dg1}
\end{figure}
\par
We take a general covariant gauge. In the gauge the gluon propagator
is given by:
\begin{equation}
  \frac{-i }{q^2 + i\varepsilon} \left ( g^{\mu\nu} - \alpha \frac{q^\mu q^\nu}{q^2 + i\varepsilon} \right )
\end{equation}
where $\alpha$ is the gauge parameter. The Feynman gauge is obtained by taking $\alpha =0$.
At one-loop the wave function receives contributions from diagrams given in Fig. 2 and Fig. 3.
The contributions from Fig. 3 are proportional to the tree-level result. The contribution from
Fig. 2b reads:
\begin{eqnarray}
\phi\vert_{2b} &=& \int \frac{ d^4 q} {(2\pi)^4} \bar v(k_2) \gamma^+ \gamma^5 \left ( ig_s u^\mu T^a \right )
       \frac{-i} { u\cdot q -i\varepsilon} \frac{ i\gamma\cdot (k_1 -q)}{ (k_1-q)^2 +i\varepsilon}
       \left (-i g_s \gamma^\nu T^a \right )
       u(k_1)
\nonumber\\
   && \cdot \frac{-i}{q^2 +i\varepsilon} \left ( g^{\mu\nu} - \alpha \frac{q^\mu q^\nu}{q^2 + i\varepsilon} \right )
        \delta (k^+ -(k_1-q)^+) \delta^2 (\vec k_T - (\vec k_1 -\vec q)_\perp ),
\end{eqnarray}
where $q$ is the momentum carried by the gluon. At the one-loop level, the wave function
will receive contributions which depend on $\alpha$. We denote
these contributions as $\phi_\alpha$ and call them gauge parts.
From Fig. 2b the gauge part is:
\begin{equation}
\phi_\alpha \vert_{2b} = i \frac{4 \alpha g_s^2 }{3 (2\pi)^3 } \int \frac{d q^-}{(2\pi)}
   \frac{ {\rm Tr } \left [ \gamma^5 \gamma^- \gamma ^+ \gamma_5 \gamma\cdot (k_1 -q) \gamma\cdot q \right ]}
      { ((k_1-q)^2 +i\varepsilon) ( q^2 + i\varepsilon)^2}
       =i \frac{16 \alpha g_s^2 }{ (2\pi)^3 } \int \frac{d q^-}{(2\pi)}
   \frac{ \left [ 2 k_1^+ q^- -\vec k_{1\perp} \cdot  \vec q_\perp - q^2 \right ]}
      { ((k_1-q)^2 +i\varepsilon) ( q^2 + i\varepsilon )^2}.
\end{equation}
The $q^-$-integral can easily be performed by taking a contour. Then one can calculate
the convolution which contributes to $H^{(1)}$:
\begin{equation}
\phi_\alpha \vert_{2b} \otimes H^{(0)}
  = \int_0^1 dx \int d^2 k_T \frac {1} { xQ^2 + k_T^2} \phi_\alpha \vert_{2b}.
\end{equation}
This integral is divergent. The divergence comes from the region of $x \to  x_0$ and $\vec k_T \to \vec k_{1\perp}$.
With $q^+ =k_1^+ -k^+$ and $\vec q_\perp = \vec k_{1\perp}-\vec k_T$ this region corresponds to the region where the gluon momentum $q$ scales as
$q^\mu =Q( \delta^2, {\mathcal O}(1), \delta, \delta)$ with  $\delta$ approaching to zero.
Here $Q$ is a scale.
In that region $q^2$ scales as $\delta^2$ and goes to zero.
This results in a divergence that comes from the first term in Eq.(13).
Since $q^-$ is much larger than $q^+$ and $\vert \vec q_\perp \vert$, the divergence is a light-cone divergence, and
the corresponding singularity can be called as light-cone singularity.
It should be emphasized that the singularity is not an infrared(I.R.)  singularity.
By isolating the divergence one can find that the divergent part of the convolution
is proportional to the divergent integral:
\begin{equation}
 \int_0^ {x_0} d x \frac{1}{x_0 -x } = \int_0^{k_1^+} \frac{d q^+}{q^+}.
\end{equation}
The singularity comes from
the end-point at $q^+ =0$. The restriction of $q^+ < k_1^+$ is from the definition
of the wave function with $k^+ = xP^+ = k_1^+ -q^+ >0$.
Later, we will discuss the appearance of the singularity in more detail.
\par
The singularity comes from the gauge part of the gluon propagator. One can introduce a gluon mass $\lambda_L$
to regularize the light-cone singularity. With the mass $\lambda_L$ the gluon propagator reads\cite{ItZu}:
\begin{equation}
 \frac{-i}{q^2 -\lambda_L^2 +i\varepsilon} \left [ g^{\mu\nu}
  - \alpha \frac{q^\mu q^\nu}{ q^2 -(1-\alpha) \lambda_L^2 + i\varepsilon } \right ].
\nonumber\\
\end{equation}
the divergent part
of $\phi_\alpha\vert_{2b}$ can be found as:
\begin{eqnarray}
\phi_\alpha \vert_{2b} &=&  -\frac {4\alpha \alpha_s }{\pi^2} \theta (x_0-x) \frac{ k^2_{1\perp}}
   {  k_1^+ \left [ q^2_\perp + y_q k^2_{1\perp} +\lambda_L^2 \right ]
   \left [ q^2_\perp + y_q k^2_{1\perp} +\bar \alpha \lambda_L^2 \right ] } +{\rm finite\ terms},
\nonumber\\
    \bar \alpha &=& 1-\alpha, \ \ \ \ \  y_q = \frac{q^+}{k_1^+}=\frac{x_0-x}{x_0}.
\end{eqnarray}
With the gluon mass we obtain the divergent part of the convolution:
\begin{equation}
\phi_\alpha \vert_{2b} \otimes H^{(0)} =
    \frac{4\alpha \alpha_s}{P^+ \pi} \frac { \ln \lambda_L^2 } { x_0 Q^2 + k_{1\perp}^2} + {\rm finite\ terms}.
\end{equation}
It should be pointed out that the contour integral of $q^-$ may cause some problems at first glance,
because the integral becomes singular in the region of $q^+\sim 0$. However, once the singularity
is regularized as in the above, the contour integral is well-defined. One can also use dimensional
regularization to regularize the singularity as the pole of $d-4$.
\par
In the convolution of Fig. 2c and Fig. 2a there are similar singularities.
Working out the singular contributions  we have for the divergent parts:
\begin{eqnarray}
\phi_\alpha \vert_{2c} \otimes H^{(0)} &=&
   \frac{4\alpha \alpha_s}{P^+ \pi}  \frac {\ln \lambda_L^2 } { x_0 Q^2 + k_{1\perp}^2} + {\rm finite\ terms},
\nonumber\\
\phi_\alpha \vert_{2a} \otimes H^{(0)} &=&
   -\frac{4\alpha \alpha_s}{P^+ \pi}  \frac {\ln \lambda_L^2 } { x_0 Q^2 + k_{1\perp}^2} + {\rm finite\ terms},
\end{eqnarray}
We note that the sum of the three parts is still divergent.
The gauge part of the contribution of Fig. 2d reads:
\begin{equation}
\phi_\alpha \vert_{2d} = i\alpha \frac{ 16 g_s^2}{(2\pi)^3} \int \frac{ dq^-} {2\pi} \frac{1}{(q^2 + i\varepsilon)^2},
\end{equation}
where $q$ is the momentum carried by the gluon.
When convoluted with a test function, one can see that this part only contributes in the region
$q^+ \sim 0$. For $q^+ \neq 0$ one can perform the $q^-$-integral and get a result of zero.
Therefore, we have:
\begin{equation}
\phi_\alpha \vert_{2d} = i\alpha \frac{ 16 g_s^2}{(2\pi)^3} \delta (x-x_0)
 \int_{k_1^+ -P^+}^{k_1^+} \frac{ d q^+}{P^+} \int \frac{ dq^-} {2\pi} \frac{1}{(q^2 + i\varepsilon)^2}.
\end{equation}
Again it is divergent. The divergence is an I.R. singularity. We regularize the singularity
with a finite gluon mass $\lambda_I$, and obtain the divergent part of the convolution:
\begin{equation}
\phi \vert_{2d} \otimes H^{(0)}
  =  \frac{4 \alpha_s}{P^+ \pi}\left ( 2 +\alpha \right ) \frac { \ln \lambda_I^2 } { x_0 Q^2 + k_{1\perp}^2}
   +{\rm finite\ terms},
\end{equation}
where we also give the singular contribution which does not depend on $\alpha$. In the above convolutions
the finite terms are also free from U.V. singularities, i.e., they do not depend on the renormalization
scale $\mu$.
\par
\begin{figure}[hbt]
\begin{center}
\includegraphics[width=8cm]{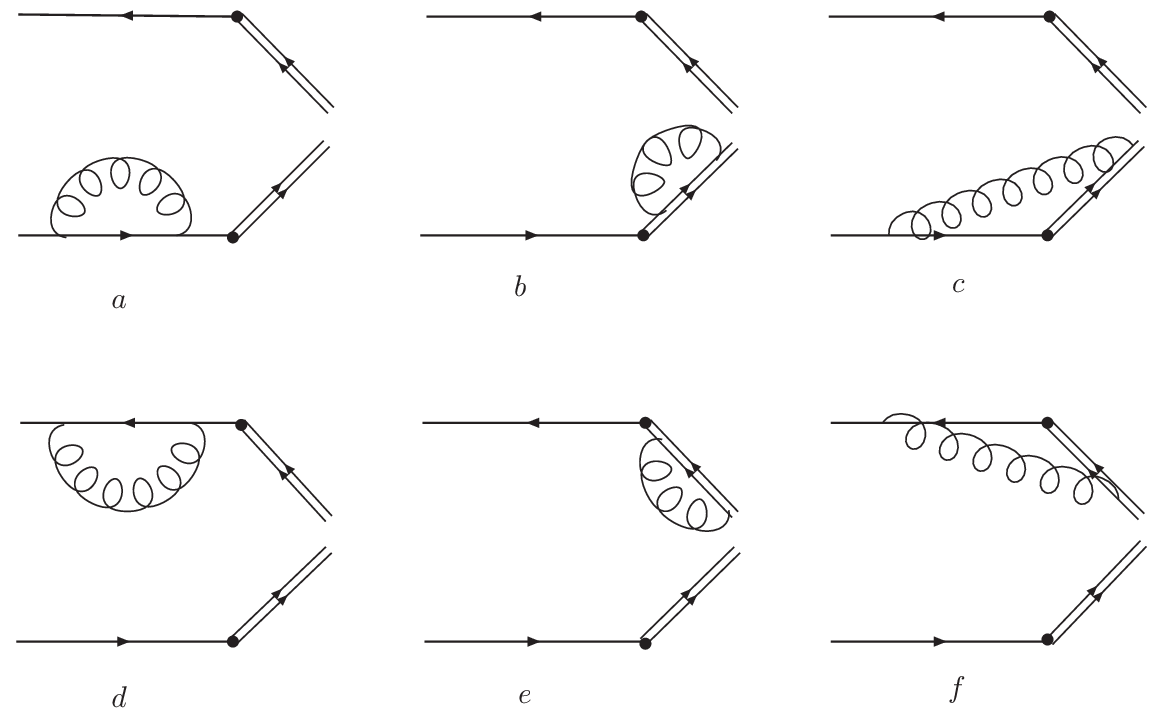}
\end{center}
\caption{The one-loop diagrams of the wave function. }
\label{Feynman-dg1}
\end{figure}
\par
Now we turn to the contributions from Fig. 3. The contributions from Fig. 3a and Fig. 3d will not
contribute to the hard part, because the same is also contained in the form factor.
The contributions from other diagrams are:
\begin{eqnarray}
\phi\vert_{3b} &=&  \phi\vert_{3e} = -\frac{2  \alpha_s}{ \pi P^+} \delta (x-x_0)
          \delta^2 (\vec k_T -\vec k_{1\perp}) \left [ \left (2+\alpha \right ) \ln \frac{\lambda^2_I}{\mu^2}
           -(1 -\alpha )
           \ln (1-\alpha )-\alpha \right ] ,
\nonumber\\
\phi\vert_{3c} &=&  \phi\vert_{3f} =-\frac{4 \alpha \alpha_s}{ \pi P^+} \delta (x-x_0)
          \delta^2 (\vec k_T -\vec k_{1\perp}) \left [ \ln \frac{\mu^2}{k_T^2} +1 \right ]+ \cdots,
\end{eqnarray}
where $\cdots$ denote contributions which do not depend on the gauge parameter $\alpha$.
In the work of \cite{NLi}
the contributions from gluon exchange between gauge links, i.e., those from Fig. 2d, Fig. 3b and Fig. 3e,
are simply neglected with the argument that these contributions or diagrams do not correspond
to any part of the form factor. It is true that  there are no corresponding parts in
the form factor $F^{(1)}$, but the contributions
from Fig. 2d, Fig. 3b and Fig. 3e  exist  by the definition of the wave function in Eq.(2).
The existence
of these contributions makes the wave function with on-shell partons
gauge invariant. Without them this wave function
is not gauge invariant. This can be shown with the general covariant gauge.
We note that  the sum of the contributions from the three diagrams to $H^{(1)}$
is free from I.R. singularities. However, the situation can become complicated
beyond the one-loop level where the exchange of gluons between gauge links and the exchange
of gluons between quark lines and gauge links can exist simultaneously, the cancellation
of the I.R. singularities in this case can be problematic. A better way to deal the problem is
to subtract these I.R. singularities
by introducing a soft factor as shown in ref. \cite{MW1,MW2}.
It should be noted that the contributions from Fig. 3 determine the $\mu$-dependence of
the wave function. This $\mu$-dependence must be gauge invariant in order to make sure that
the $\mu$-dependence of $H$ is gauge invariant, because the form factor does not depend on
$\mu$. Neglecting the contributions from Fig. 3b and Fig. 3e, as in ref. \cite{NLi}, causes
the $\mu$-dependence of the wave function to not be gauge invariant. This already implies
that the hard part determined in ref. \cite{NLi} is not gauge invariant.
\par
It should be emphasized that the contributions from Fig. 3c and Fig. 3f do not have the light-cone
singularity by our explicit calculation. For our conclusion presented
in this letter, it is crucial to understand the absence of the singularity here. Part of the contribution
of Fig. 3c is proportional to the integral
\begin{equation}
  I^\mu = \int \frac {d^4 q}{(2\pi)^4} \frac{q^\mu}{(q^2 +i\varepsilon)^2 ( (k_1-q)^2 +i\varepsilon)}.
\end{equation}
The light-cone singularity can appear in the component $I^-$ of $I^\mu$. From the Lorentz covariance
one has $I^\mu \propto k_1^\mu$. This leads to $I^- =0$ because $k^-_1 =0$.
Therefore, the light-cone singularity does not exist in the contribution of Fig. 3c.
The nonexistence can also be understood in another way, which is important
for the later discussion of the form factor. By power counting in the
momentum region $q^\mu =Q( \delta^2, {\mathcal O}(1), \delta, \delta)$ one can find
the leading contribution of $I^-$ which can have  the singularity:
\begin{equation}
  I^-_{L} = \int \frac {d^4 q}{ (2\pi)^4} \frac{ q^-}{(q^2 +i\varepsilon)^2 ( k_1^2- 2 k_1^+ q^- +i\varepsilon)}.
\end{equation}
There is an ambiguity in the order of the integration.
If we perform the $q^+$-integration first by taking a contour or by integrating directly, one simply
gets $I^-_L =0$. If we first perform the $q^-$-integral and subsequently the $\vec q_\perp$-integral,
we get the result:
\begin{equation}
 I^-_{L} \propto \int_0^\infty d q^+ \frac{1}{q^+}.
\end{equation}
However, the integral can be correctly evaluated by writing the denominator in a
covariant form. We note that $k_1^- =0$ and introduce a vector $\tilde k_1^\mu = (k_1^+, 0,0,0)$,
or $\tilde k_1^\mu = (k_1^0, 0,0,k_3^0)$ in the usual coordinate system.
With the introduced vector we have $2 k_1^+ q^- = 2 \tilde k_1 \cdot q$.
The integral $I^-_{L}$ can be obtained as a component of the vector:
\begin{equation}
I^\mu_{L} = \int \frac {d^4 q}{ (2\pi)^4} \frac{ q^\mu}{(q^2 +i\varepsilon)^2 ( k_1^2- 2 \tilde k_1\cdot q +i\varepsilon)}.
\end{equation}
With the standard method of loop-integrals or Lorentz covariance we immediately get
$I^\mu_{L} \propto \tilde k_1^\mu$, and therefore $I_L^- =0$.
This indicates that $I^-$ does not contain the light-cone singularity as expected.
It is clear that the result of $I^-_L =0$ depends on the fact that the $q^+$-integration is unrestricted
and the integral can be performed in a covariant way.
In the convolution of $H^{(0)}$ with the contribution from Fig. 2b the corresponding integration region
of $q^+$ is restricted by definition, hence the light-cone singularity appears.
\par
It may be interesting to study in more detail why $I^-$ related to Fig. 3c is finite and why the similar integral
related to the convolution $\phi_\alpha \vert_{2b} \otimes H^{(0)}$, where the integrand of $I^-$ appears
as a part of the integrand in the convolution, is singular. For this purpose we use the dimensional
regularization with the transverse momentum in $2-\epsilon$ space. One can perform the integral $I^-$
in the light-cone coordinate system. In this case one not only meets the light-cone singularity mentioned
before, but also the light-cone singularity in the momentum region of $q^\mu =Q({\mathcal O}(1), \Lambda^2,
\Lambda, \Lambda)$ with $\Lambda \to \infty$. The two singularities are canceled, and in the limit
$\epsilon \to 0$ one finds $I^- =0$ in agreement with the argument from the covariance. For the convolution
the integrand is a product of $H^{(0)}$ and the integrand of $I^-$:
\begin{eqnarray}
\phi_\alpha \vert_{2b} \otimes H^{(0)} &\propto& \int^{x_0 P^+}_0 d q^+ \int \frac{ d q^- d^{2-\epsilon} q_\perp}
    { (2\pi)^3 }
  \left [ \frac{1}{ (x_0 P^+ -q^+) (Q^2/P^+) + (\vec k_{1\perp} - \vec q_\perp)^2} \right ]
\nonumber\\
  && \cdot \left [ \frac{q^-}{(q^2 +i\varepsilon)^2 ( (k_1-q)^2 +i\varepsilon)} \right ] + {\rm finite\ terms},
\end{eqnarray}
where the $[ \cdots ]$ in the first line is $H^{(0)}$, and the $[ \cdots ]$ in the
second line is the integrand of $I^-$. Because the $q_\perp$-dependence in
$H^{(0)}$, the integral is finite in the momentum region
of $q^\mu =Q({\mathcal O}(1), \Lambda^2, \Lambda, \Lambda)$ with $\Lambda \to \infty$.
But the integral is divergent in the momentum region of $q^\mu =Q(\delta^2,{\mathcal O}(1),
\delta, \delta)$ with $\delta \to 0 $ as found before. Unlike the integral $I^-$, the divergence is not canceled
here. In Eq. (18) the divergence is regularized with the gluon mass.
Detailed results with the dimensional regularization can be found in ref. \cite{repl}.

\par
\begin{figure}[hbt]
\begin{center}
\includegraphics[width=11cm]{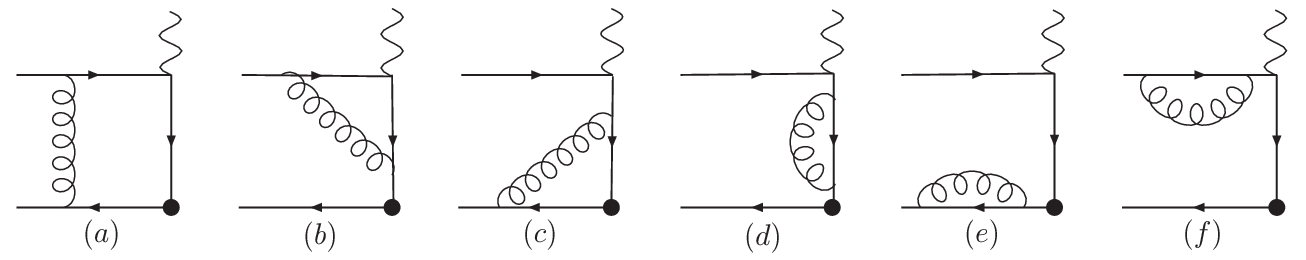}
\end{center}
\caption{The one-loop diagrams of the form factor. }
\label{Feynman-dg1}
\end{figure}
\par
The one-loop correction of the form factor comes from the diagrams given in Fig. 4.
These diagrams are ordinary Feynman diagrams of QCD Green functions.
They
do not have the light-cone singularity in the contributions depending on $\alpha$.
Taking the gauge part of the contribution
of Fig. 4b as an example, the possible light-cone
singularity is contained in the component $J^{--}$ of the tensor
$J^{\mu\nu}$ which is:
\begin{equation}
J^{\mu\nu} = \int \frac{ d^4 q}{(2\pi)^4} \frac {q^\mu q^\nu }{ (q^2 +i\varepsilon)^2}
 \frac{1}{(k_1 -q)^2 + i\varepsilon} \frac{1}{(k_1 -p -q)^2 +i\varepsilon},
\end{equation}
where $p$ is the momentum of the photon.
By the momentum scaling of $q^\mu$ we find the leading contribution
of $J^{--}$ which can have the singularity:
\begin{equation}
J^{--}_L = \int \frac{ d^4 q}{(2\pi)^4} \frac {q^- q^- }{ (q^2 +i\varepsilon)^2}
 \frac{1} { k_1^2 - 2k_1^+ q^- +i\varepsilon}\frac{1}{(k_1 -p)^2 -2 k_1^+ q^-  +i\varepsilon}.
\end{equation}
With the argument given before, $J^{--}_L$ is zero. Therefore, $J^{\mu\nu}$ does not contain
the light-cone singularity. This has been also checked for $J^{\mu\nu}$
with the standard method for loop integrals and the absence of the singularity has been confirmed.
With this fact, the contribution from Fig. 4b is free from light –cone singularities.
With the same argument one also finds that the contributions from Fig. 4a, Fig. 4c and Fig. 4d
are free from light-cone singularities. This is also confirmed by calculating the
relevant integrals with the standard method for loop integrals.
This is in agreement with the expectation that ordinary Feynman diagrams of QCD Green functions
have only U.V.-, I.R.- and collinear singularities.
\par
With the above results
the hard part $H^{(1)}$ at one-loop level will receive a contribution which contains
a light-cone singularity and the contribution depends on
the gauge parameter $\alpha$. This leads to the conclusion
that the $k_T$-factorization with off-shell partons given in the form of Eq.(1) does not hold at one-loop level
and the factorization is {\it not} gauge invariant. Since these singularities
come from the wave function and the scattering amplitude of off-shell partons do not
contain the same light-cone singularities, the $k_T$-factorization is violated not only
in the case studied here but also in all cases.
\par
In applications of the $k_T$-factorization for $B$-decays
one can also introduce the $k_T$-dependent wave function for $B$-mesons,
where one replaces in Eq.(2) $\pi^0$ with a $B$-meson and the anti-quark field $\bar q$
with the effective field of $b$-quark. At one-loop the wave function of a $B$-meson will
receive contributions from Fig. 2 and Fig. 3, where the anti-quark line is replaced
with a gauge link along the direction $v$, where $v$ is the four velocity
of the $B$-meson. This gauge link stands for the $\bar b$-quark.
The hard part will receive contributions of scattering amplitudes of off-shell partons, which correspond
to the form factor in Eq.(10), and contributions from the wave function of the $B$-meson.
The contributions of scattering of off-shell partons do not contain
light-cone singularities, while the wave function of the $B$-meson
contains these singularities, according to our results.
It is clear that the hard part is not gauge invariant and contains light-cone singularities
in a general covariant gauge. Therefore the $k_T$-factorization also does not hold
in exclusive $B$-decays if one takes off-shell partons to perform the factorization.
Detailed examples and results will be given in a future work.
\par
It is possible to perform a factorization by taking transverse momenta
of partons into account, where partons are on-shell.
This has been studied in ref. \cite{MW1,MW2} for exclusive processes in which
only one hadron is involved. It has been shown that an additional soft factor $S$
is needed. In the case studied explicitly here, the factorization
reads\cite{MW2}:
\begin{equation}
 F(Q^2 ) \sim \phi \otimes  S \otimes  H \left ( 1 + {\mathcal O} (Q^{-2}) \right )
\end{equation}
where $\phi$ is defined in Eq.(2) and the soft factor $S$ can be found in ref. \cite{MW1,MW2}.
The hard part $H$ in the above is calculable with perturbation theory. To distinguish
the $k_T$-factorization with off-shell partons, the factorization in the above is called
Transverse-Momentum-Dependent(TMD) factorization. With TMD factorization
the Sudakov logarithms can also be re-summed. The re-summation can even be done
within the collinear factorization with the usage of gauge links\cite{FMW}.
\par
To summarize: We have studied the $k_T$-factorization for the case $\pi\gamma^* \to \gamma$
at one-loop in a general covariant gauge. In the $k_T$-factorization
the nontrivial $k_T$-dependence
of the hard part is obtained by taking off-shell partons. With
off-shell partons we show that the hard part at one-loop contains a light-cone singularity
in the general covariant gauge. The singularity is absent in Feynman gauge.
This indicates that the $k_T$-factorization is not gauge invariant.
These singular contributions violate the $k_T$-factorization.
The singular contributions come from the
one-loop correction of the wave functions and they are not related
to a special process. Based on this fact one can conclude
that the $k_T$-factorization for any exclusive process is not gauge invariant and
does not hold. The $k_T$-factorization has been widely used for exclusive
$B$-decays by introducing the wave function of $B$-mesons.
Our results indicate that the wave function of $B$-mesons with off-shell partons
also contain light-cone singularities. This results in
the hard part of the $k_T$-factorization for exclusive $B$-decays
containing these singularities. Therefore, the $k_T$-factorization
for exclusive $B$-decays is violated as well.
\par\vskip20pt
\par\noindent
{\bf\large Acknowledgments}
\par
We thank Prof. M. Yu and Prof. H.-n. Li for interesting discussions.
This work is supported by National Nature Science Foundation of P.R. China(No. 10721063,10575126,10747140).
\par



\vskip20pt

\begin{thebibliography}{99}

\bibitem{BL} G.P. Lepage and S.J. Brodsky, Phys. Rev. D22 (1980) 2157.

\bibitem{CZrep} V.L. Chernyak and A. R. Zhitnitsky, Phys. Rept.  {\bf 112}
(1984) 173.

\bibitem{LS} H.-n. Li and G. Sterman, Nucl. Phys. B381 (1992) 129.

\bibitem{KTB1} H.-n. Li and H.L. Yu, Phys. Rev. Lett. {\bf 74} (1995) 4388;
Phys. Lett. B353 (1995) 301; Phys. Rev. D53 (1996) 2480,
H.-n. Li and B. Tseng, Phys. Rev. D57 (1998) 443,
H.-n. Li and G.-L. Lin, Phys. Rev. D60 (1999) 054001,
T. Kurimoto, H-n. Li, and A.I. Sanda, Phys. Rev D65 (2002) 014007; Phys. Rev. D67 (2003) 054028,
H.-n. Li, Phys.Lett. B622 (2005) 63,
H.-n. Li, S. Mishima and A.I. Sanda, Phys.Rev. D72 (2005) 114005,
Y.-Y. Charng, T. Kurimoto and  H.-nan Li, Phys.Rev.D74 (2006) 074024.

\bibitem{KTB2}
A. Ali, G. Kramer, Y. Li, C.-D. Lu, Y.-L. Shen, W. Wang and  Y.-M. Wang, Phys.Rev. D76 (2007) 074018,
C.-D. Lu, W. Wang and Y.-M. Wang, Phys. Rev. D75 (2007) 094020,
Y. Li and C.-D. Lu, Phys.Rev.D74:097502,2006,
X.-Q. Yu, Y. Li and C.-D. L\"u, Phys.Rev. D73 (2006) 017501,
C.-D. Lu, M. Matsumori, A.I. Sanda and M.-Z. Yang, Phys. Rev. D72 (2005) 094005,
G.-L. Song and C.-D. Lu, Phys.Rev. D70 (2004) 034006,
C.-D. Lu, K. Ukai and M.-Z. Yang, Phys. Rev. D63 (2001) 074009,
D.-Q. Guo, X.-F. Chen and Z.-J. Xiao, Phys. Rev. D75 (2007) 054033,
H.-s. Wang, X. Liu, Z.-J Xiao, L.-B. Guo and C.-D. Lu, Nucl. Phys. B738 (2006) 243,
X.-G. He, T. Li, X.-Q. Li and Y.-M. Wang, Phys.Rev. D75 (2007) 034011,
X.-Q. Li, X. Liu and Y.-M. Wang, Phys. Rev. D74 (2006) 114029,
T. Kurimoto, Phys. Rev. D74 (2006) 014027,
Y.Y. Keum, M. Matsumori and A.I. Sanda, Phys. Rev. D72 (2005) 014013,
S. Mishima and A.I. Sanda, Phys. Rev. D69 (2004) 054005,
Z.-T. Wei and M.-Z. Yang, Nucl. Phys. B642 (2002) 263, Phys. Rev. D67 (2003)094013.


\bibitem{NLi} S. Nandi and H.-n. Li, Phys. Rev. D76 (2007) 034008.

\bibitem{LiLi} H.-n. Li and H.-S. Liao, Phys. Rev. D70 (2004) 074030.

\bibitem{MW1} J.P. Ma and Q. Wang, Phys. Lett. B613 (2005) 39, JHEP 0601 (2006) 067.

\bibitem{MW2} J.P. Ma and Q. Wang, Phys. Lett. B642 (2006) 232, Phys. Rev. D75 (2007) 014014.

\bibitem{ItZu} C. Itzykson and J. Zuber, {\it Quantum Field Theory}, New York, McGraw-Hill (1980).

\bibitem{repl} F. Feng, J.P. Ma and Q. Wang, e-Print: arXiv:0808.4017 [hep-ph].

\bibitem{FMW} F. Feng, J.P. Ma and Q. Wang, JHEP 0706 (2007) 039.









\end{thebibliography}
\end{document}